\documentclass[10pt,twocolumn,letterpaper]{article}

\usepackage[pagenumbers]{cvpr} 

\usepackage{graphicx}
\usepackage{amsmath}
\usepackage{amssymb}
\usepackage{booktabs}
\usepackage{graphicx}
\usepackage{amsmath,amssymb,amsfonts}
\usepackage{multirow}
\usepackage{xcolor}
\usepackage{breqn}
\usepackage{url}
\usepackage{ulem}
\usepackage{appendix}
\usepackage{url}

\usepackage[pagebackref,breaklinks,colorlinks]{hyperref}

\usepackage[capitalize]{cleveref}
\crefname{section}{Sec.}{Secs.}
\Crefname{section}{Section}{Sections}
\Crefname{table}{Table}{Tables}
\crefname{table}{Tab.}{Tabs.}

\def\cvprPaperID{*****} 
\def\confName{CVPR}
\def\confYear{2023}

\begin{document}

\title{A Generalized Surface Loss for Reducing the Hausdorff Distance in Medical Imaging Segmentation}

\author{Adrian Celaya\\
Rice University\\
Houston, TX\\
{\tt\small aecelaya@rice.edu}
\and
Beatrice Riviere\\
Rice University\\
Houston, TX\\
{\tt\small riviere@rice.edu}
\and
David Fuentes\\
The University of Texas MD Anderson Cancer Center\\
Houston, TX\\
{\tt\small dtfuentes@mdanderson.org}
}
\maketitle

\begin{abstract}
Within medical imaging segmentation, the Dice coefficient and Hausdorff-based metrics are standard measures of success for deep learning models. However, modern loss functions for medical image segmentation often only consider the Dice coefficient or similar region-based metrics during training. As a result, segmentation architectures trained over such loss functions run the risk of achieving high accuracy for the Dice coefficient but low accuracy for Hausdorff-based metrics. Low accuracy on Hausdorff-based metrics can be problematic for applications such as tumor segmentation, where such benchmarks are crucial. For example, high Dice scores accompanied by significant Hausdorff errors could indicate that the predictions fail to detect small tumors. We propose the Generalized Surface Loss function, a novel loss function to minimize Hausdorff-based metrics with more desirable numerical properties than current methods and with weighting terms for class imbalance. Our loss function outperforms other losses when tested on the LiTS and BraTS datasets using the state-of-the-art nnUNet architecture. These results suggest we can improve medical imaging segmentation accuracy with our novel loss function.
\end{abstract}

\section{Introduction}
\label{sec:intro}
Deep learning has become a popular framework in medical image analysis, automating and standardizing essential tasks such as segmenting regions of interest. This process is crucial for computer-assisted diagnosis, intervention, and therapy \cite{intro-2}. Despite its importance, manual image segmentation can be a tedious and time-consuming task with variable results among users \cite{intro-1, intro-3}. Fully automated segmentation offers a solution, reducing time and producing more consistent results \cite{intro-1, intro-3}. Convolutional neural networks (CNNs) have achieved notable success in segmentation tasks, including labeling tumors and anatomical structures \cite{unet, unet-3d, nnunet}.

\begin{figure}
    \centering
    \includegraphics[width=0.8\columnwidth]{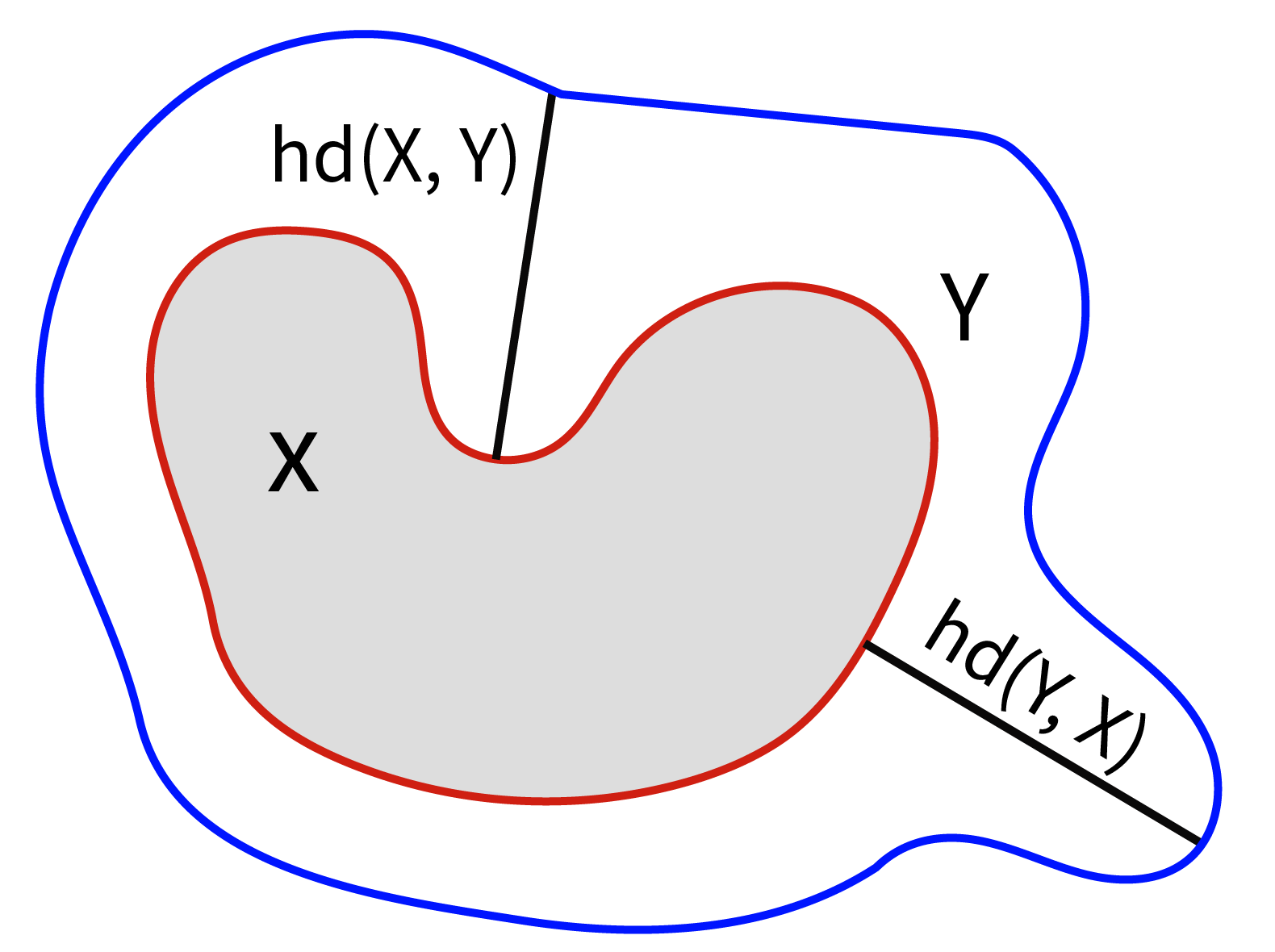}
    \caption{Illustration of the Hausdorff distance between two sets of points $X$ and $Y$. \label{fig:hd}}
\end{figure}

In medical imaging segmentation, the performance of automatic methods is usually evaluated using common metrics like the Dice similarity coefficient, average surface distance (ASD), and the Hausdorff distance (HD) \cite{crum2006generalized, taha2015metrics}. The HD is beneficial among these metrics because it indicates the largest segmentation error. As sketched out in Figure \ref{fig:hd}, for two sets of points $X$ and $Y$, the one-sided HD from $X$ to $Y$ is given by
\begin{align}
    \text{hd}(X, Y) = \max_{x \in X} \min_{y \in Y} ||x - y||_2.
\end{align}
The one-sided HD is not a true distance metric since it is not commutative (i.e., $\text{hd}(X, Y) \neq \text{hd}(Y, X)$). To address this, we consider the bidirectional or total Hausdorff distance, which is given by
\begin{align} \label{eqn:hd}
    \text{HD}(X, Y) = \max \{ \text{hd}(X, Y), \text{hd}(Y, X) \}.
\end{align}
In (\ref{eqn:hd}), we use the Euclidean distance, but any other distance metric can be used. Intuitively, the HD is the longest distance from one point in a set to the closest point in the other.

Although the HD and other similar metrics like the ASD are widely used for evaluating medical imaging segmentation models, many current loss functions for medical image segmentation only consider the Dice coefficient or similar region-based metrics during training \cite{loss-survey, salehi2017tversky, abraham2019novel, diceloss, bertels2019optimizing}. This approach runs the risk of achieving high accuracy for the Dice coefficient but low accuracy for Hausdorff-based metrics \cite{bl, hausloss, taha2015metrics}. This low accuracy is particularly problematic for applications such as tumor segmentation, where Hausdorff-based metrics are crucial for evaluating segmentation accuracy \cite{hd-importance}. For example, high Dice scores accompanied by significant Hausdorff errors could indicate that the predictions fail to detect small tumors. As a result, we propose the Generalized Surface Loss function, a novel loss function to minimize the Hausdorff distance (HD). This loss function incorporates a normalization scheme to give it more desirable numerical properties than current methods and uses weighting terms to address class imbalance.

\subsection{Previous Work}
\label{sec:prevwork}
Over the last several years, multiple loss functions have been developed for medical image segmentation models. Broadly, these loss functions fall into two categories: region and boundary-based loss functions. We briefly survey these loss functions below.

\subsubsection{Region-Based Losses}
\paragraph{Dice Loss} - The Dice coefficient is a widely used metric in computer vision tasks to calculate the global measure of overlap between two binary sets \cite{dice}. Introduced by Milletari et al. \cite{vnet}, the Dice Loss (DL) function directly incorporates this metric into the formulation as follows:
\begin{align} \label{eqn:dice}
    \mathcal{L}_{dice} = 1 - \frac{1}{C}\sum_{k=1}^C \frac{2\sum_{i=1}^N T_i^k P_i^k}{\sum_{i=1}^N \left(T_i^k\right)^2 + \sum_{i=1}^N \left(P_i^k\right)^2},
\end{align}
where $N$ denotes the total number of pixels (or voxels in the 3D case), $C$ denotes the number of segmentation classes, $P_i^k \in [0, 1]$ is the $i$-th voxel in the $k$-th class of the predicted segmentation mask, and $T_i^k \in \{0, 1\}$ is the same for the ground truth. It is also common for the DL to be used in conjunction with the cross entropy loss function \cite{nnunet, loss-survey, zhu2022segmentation, taghanaki2019combo}. This composite loss (which we call the Dice-CE loss) adds a cross-entropy term to (\ref{eqn:dice}) and is given by 
\begin{align}
    \mathcal{L}_{dice-ce} = \mathcal{L}_{dice} - \frac{1}{CN} \sum_{k=1}^C \sum_{i=1}^N T_i^k \log\left(P_i^k\right)
\end{align}

The use of DL (and Dice-CE loss) in the training of deep learning models for medical imaging segmentation tasks is widespread \cite{nnunet, loss-survey, actor2020identification, ma2021loss, zhu2019anatomynet, bertels2019optimizing}. This is because the Dice coefficient is a widely recognized metric for measuring the global overlap between the predicted and ground truth segmentation masks. However, the DL is a region-based loss that does not take into account the HD during training. As a result, models trained using the DL function can achieve high accuracy with respect to the Dice coefficient, but exhibit poor accuracy with respect to HD-based metrics \cite{asaturyan2019advancing}. This low accuracy is particularly problematic for tasks such as tumor segmentation, where HD-based metrics are crucial to evaluate the accuracy of segmentation \cite{hd-importance}.

\paragraph{Generalized Dice Loss} - Sudre et al. proposed the Generalized Dice Loss (GDL) \cite{diceloss} by introducing a weighting term to the DL function presented above. In their work, the addition of these weight terms produced better Dice scores for highly imbalanced segmentation problems. The GDL loss is given by the following:
\begin{align}
    \mathcal{L}_{gdl} = 1 - 2\frac{\sum_{k=1}^{C} w_k \sum_{i=1}^{N} T_i^k P_i^k}{\sum_{k=1}^{C} w_k \sum_{i=1}^{N} \left(\left(T_i^k\right)^2 + \left(P_i^k\right)^2\right)},
\end{align}
where the term $w_k$ is the weighting term for the $k$-th class and is given by 
\begin{align}
    w_k = \frac{1}{ \left(\sum_{i=1}^{N} T_i^k \right)^2}.
\end{align}

The intuition behind $w_k$ is that the contribution of each labeled class is the inverse of its area (or volume in the 3D case). Hence, this weighting can help reduce the well-known correlation between region size and the Dice score \cite{diceloss, liu2023we, wong20183d}. However, like the DL, the GDL is a region-based loss and does not consider the HD, and can result in poor accuracy for HD-based metrics. Additionally, the weights $w_k$ change for each training example, which makes the optimization process more challenging, as the problem is altered for every batch.


\subsubsection{Boundary-Based Losses}
\begin{figure}
    \centering
    \includegraphics[width=\columnwidth]{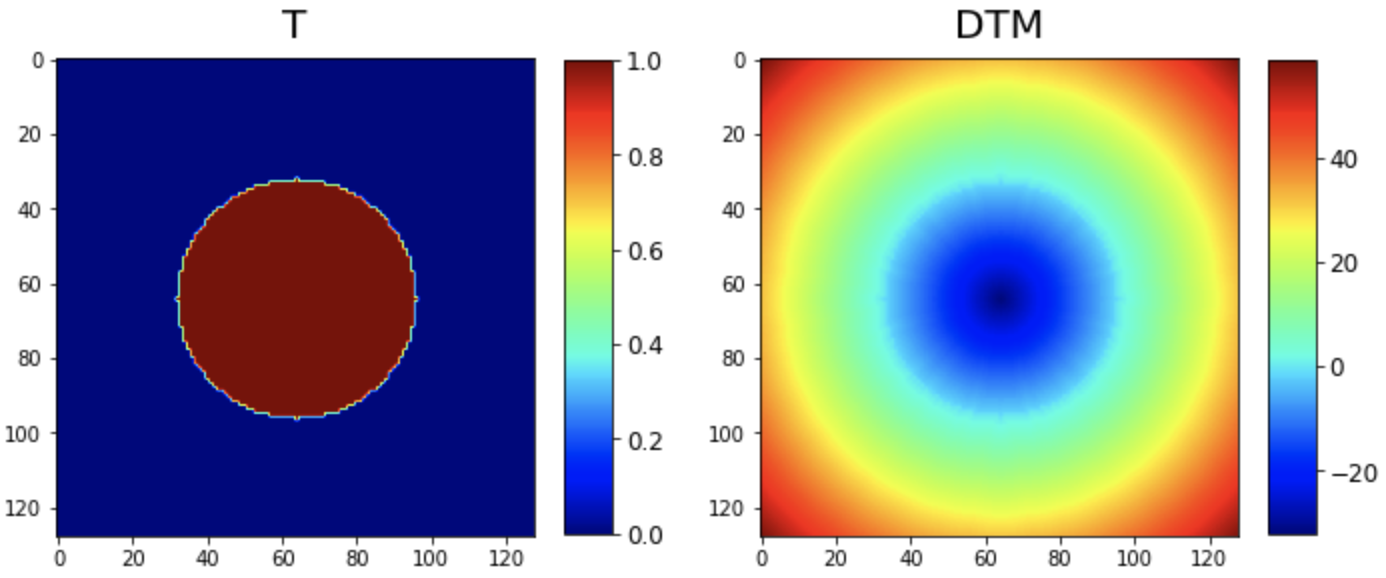}
    \caption{(Left) Example segmentation. (Right) DTM for example segmentation. Here, the values in the DTM are positive on the exterior, zero on the boundary, or negative in the object's interior.\label{fig:dtm}}
\end{figure}

\paragraph{Hausdorff Loss} - Karimi et al. introduced a Hausdorff Loss (HL) that incorporates an estimation of the HD in the loss function via Distance Transform Maps (DTM) \cite{bl}. A DTM represents the distance between each voxel and the closest boundary or edge of an object. The values in a DTM are positive on the exterior, zero on the boundary, or negative in the interior of the object. Figure \ref{fig:dtm} illustrates an example segmentation image and its corresponding DTM.

The HL is given by the following:
\begin{align} \label{eqn:hl}
    \mathcal{L}_{haus} = \frac{1}{CN} \sum_{k=1}^{C} \sum_{i=1}^N (T_i^k - P_i^k)^2\left(\left(D_i^k\right)^2 + \left(\bar{D}_i^k\right)^2\right),
\end{align}
where $D_i^k$ is the $i$-th voxel in the DTM for the $k$-th class in the ground truth, and where $\bar{D}_i^k$ is the same for the prediction. This function is used in a weighted combination with a region-based loss $\mathcal{L}_{region}$ so that the overall loss function is given by
\begin{align} \label{eqn:scheduled-loss}
    \mathcal{L} = \alpha \mathcal{L}_{region} + (1 - \alpha) \mathcal{L}_{haus}
\end{align}
where the scalar $\alpha=1$ in the first epoch and decreases linearly until $\alpha=0$. While the HL does consider the HD, it requires the re-computation of the DTM for each predicted segmentation at each epoch. It also requires the storage of the whole volume in memory since one cannot compute the DTM for a subset of the volume. Although the DTM can be computed in linear time \cite{maurer2003linear}, reconstructing the DTM of the whole volume for each training batch results in a significant slowdown in training \cite{hausloss}. These two factors make the HL computationally expensive and difficult to incorporate into current patch-based medical imaging segmentation pipelines like nnUNet, MONAI, or MIST \cite{nnunet, cardoso2022monai, celaya2022pocketnet}. To address these issues, Karimi et al. introduce a one-sided version of (\ref{eqn:hl}), which does not consider the DTM of the prediction. The HL will refer to this one-sided version for the remainder of this work.

\paragraph{Boundary Loss} - Kervadec et al. proposed a Boundary Loss (BL) function, which considers the measure of distance to the boundary of a region \cite{bl}. Like the HL function, the BL uses a DTM to measure the distance to the boundary of the ground truth image. The BL loss is given by the following:
\begin{align}
    \mathcal{L}_{bl} = \frac{1}{CN} \sum_{k=1}^{C} \sum_{i=1}^N D_i^kP_i^k.
\end{align}
Also like the HL, the BL uses a weighted combination with a region-based loss so that the overall loss is given by an expression similar to (\ref{eqn:scheduled-loss}), but with $\mathcal{L}_{bl}$. Again, $\alpha = 1$ at the start of training and decreases linearly until $\alpha = 0$ in the final epoch. 

Intuitively, the BL function measures the average distance of each voxel in the predicted segmentation mask to the boundary of the ground truth mask. Unlike the HL, the BL does not require the re-computation of DTMs at each epoch, making it comparatively inexpensive from a computational point of view. However, region-based loss functions such as DL and GDL are bounded, achieving values in the interval [0,1]. On the other hand, the BL function is unbounded since $D_i^k \in (-\infty, \infty)$. This unboundedness can result in large positive or negative loss values and minimums that depend on individual image sizes and voxel spacings, making optimization more difficult. This same issue can also dominate the total loss, causing the contribution of the region-based loss to become negligible during training.

\section{Materials and Methods}
\label{sec:methods}
\subsection{Generalized Surface Loss}
We seek a boundary-based loss function that is 1) bounded in the interval $[0, 1]$ to ensure that its contribution to the overall loss function does not dominate the region-based loss in (\ref{eqn:scheduled-loss}), 2) is computationally tractable like the BL, and 3) is weighted by pre-computed weighting terms to handle class imbalance while not changing the optimization problem for each batch like the GDL. Now, consider a ``worst'' case prediction $\mathbf{1} - T$, which maximizes the value 
\begin{align} \label{eqn:worst}
    \sum_{i=1}^N \left( D_i \left( \left(1 - T_i\right) - T_i \right) \right)^2.
\end{align}
Note that a worst case is dependent on the choice of metric, but for our purposes we choose to maximize (\ref{eqn:worst}) so that for a given prediction $P$, we have 
\begin{align}
    0 \leq \frac{\sum_{i=1}^N \left( D_i \left( (1 - T_i) - P_i \right) \right)^2}{\sum_{i=1}^N \left( D_i \left( (1 - T_i) - T_i \right) \right)^2} \leq 1.
\end{align}
Simplifying, adding weighting terms for class imbalance, and subtracting the final value from one gives us the Generalized Surface Loss (GSL):
\begin{align}
    \mathcal{L}_{gsl} = 1 - \frac{\sum_{k=1}^C w_k \sum_{i=1}^N \left( D_i^k \left( 1 - \left(T_i^k + P_i^k\right)\right)\right)^2}{\sum_{k=1}^C w_k \sum_{i=1}^N \left( D_i^k \right)^2}.
\end{align}

We pre-compute the class weights $w_k$ so that the inverse of the total number of voxels belonging to each class over the entire dataset $N_k$ is normalized by the sum of the other inverses. For the $k$-th segmentation class, $w_k$ is given by
\begin{align} \label{eqn:weights}
    w_k = \left(\frac{1}{\sum_{j=1}^{C} \frac{1}{N_j}}\right) \frac{1}{N_k}.
\end{align}
Like BL and HL, the overall loss function uses a scheme similar to (\ref{eqn:scheduled-loss}). However, in addition to a linear schedule, we also propose the use of the step and cosine functions as schedules for setting $\alpha$ at each epoch. In each case, we decrease until $\alpha = 0$ in the final epoch. To define each schedule, let $h \geq 1$ be our step length, $t$ be the current epoch, and $T$ be the total number of epochs. Then define $N_h = \lfloor T/h \rfloor$. The value of $\alpha$ at $t$ for each schedule is given by 
\begin{align}
    \alpha_{linear}(t) &= 1 - \frac{t}{T} \\
    \alpha_{step}(t) &= 1 - \frac{\lfloor t/h \rfloor}{N_h} \\
    \alpha_{cosine}(t) &= \frac{1}{2}\left(1 + \cos{\frac{\pi t}{T}} \right)
\end{align}

\subsection{Network Architecture}
We examine the effects of our proposed loss function on the widely used nnUNet architecture \cite{nnunet}. The nnUNet architecture uses an adaptive framework based on the properties of the given dataset (i.e., patch size and voxel spacing) to build a U-shaped architecture like the 3D U-Net \cite{unet, unet-3d}. As a result, the network has achieved state-of-the-art accuracy on several recent public medical imaging segmentation challenges \cite{brats-nnunet, antonelli2022medical, heller2021state}. In the context of our work, the nnUNet architecture provides a state-of-the-art baseline for our analysis.

\subsection{Data}
\label{sec:data}
We test our proposed loss function on the MICCAI Liver and Tumor Segmentation (LiTS) Challenge 2017 dataset \cite{lits} and the multi-label tumor segmentation in the MICCAI Brain Tumor Segmentation (BraTS) Challenge 2020 dataset \cite{brats1, brats2, brats3}. In the context of our work, these datasets allow us to test each loss function on two different imaging modalities (i.e., CT and MR) and have non-trivial preprocessing and segmentation tasks. We briefly describe these datasets and the preprocessing steps we take below. 

For the LiTS dataset, we perform binary liver segmentation. This dataset consists of the 131 CT scans from the MICCAI 2017 Challenge's multi-institutional training set. These scans vary significantly in the number of slices in the axial direction and voxel resolution, although all axial slices are at 512$\times$512 resolution. As a result, we use the preprocessing steps proposed by \cite{nnunet} to handle this variability. Namely, we resample each image to the median resolution of the training data in the $x$ and $y$-directions and use the 90th percentile resolution in the $z$-direction. For intensity normalization, we window each image according to the foreground voxels' 0.5 and 99.5 percentile intensity values across all of the training data. This scheme results in windowing from -17 to 201 HU. We also apply z-score normalization according to the foreground voxels' mean and standard deviation. The LiTS dataset is available for download \url{https://competitions.codalab.org/competitions/17094#learn_the_details-overview}.

The BraTS training set contains 369 multimodal scans from 19 institutions. Each set of scans includes a T1-weighted, post-contrast T1-weighted, T2-weighted, and T2 Fluid Attenuated Inversion Recovery volume along with a multi-label ground truth segmentation. The annotations include the GD-enhancing tumor (ET - label 4), the peritumoral edema (ED - label 2), and the necrotic and non-enhancing tumor core (NCR/NET - label 1). The final segmentation classes are the whole tumor (WT - labels 1, 2, and 4), tumor core (TC - labels 2 and 4), and ET. All volumes are provided at an isotropic voxel resolution of 1$\times$1$\times$1 $\text{mm}^3$, co-registered to one another, and skull stripped, with a size of 240$\times$240$\times$155. We crop each image according to the brainmask (i.e., non-zero voxels) and apply z-score intensity normalization on only non-zero voxels for preprocessing. The BraTS training dataset is available for download at \url{https://www.med.upenn.edu/cbica/brats2020/registration.html}.

\subsection{Training and Testing}
We compare the effect on segmentation performance for each loss function described in Section \ref{sec:prevwork} and \ref{sec:methods}. We select $\mathcal{L}_{region}$ for our boundary-based loss functions to be the Dice-CE loss. We use the Adam optimizer with the learning rate set to 0.0003. A five-fold cross-validation scheme is used to train and evaluate each loss function. During training, we select a batch size of two,   using a patch size of 128$\times$128$\times$128 for the BraTS dataset and 256$\times$256$\times$128 for the LiTS dataset. We apply the same random augmentation described in \cite{nnunet}. Our models are implemented in Python using PyTorch (v2.0.1) and trained on two NVIDIA Quadro RTX 8000 GPUs using the DistributedDataParallel module \cite{paszke2019pytorch, li2020pytorch}. All network weights are initialized using the default PyTorch initializers. For reproducibility, we set all random seeds to 42. All other hyperparameters are left at their default values. The code for this work is available at \url{https://anonymous.4open.science/r/gen-surf-loss-B624/README.md}.

We used three common medical imaging segmentation metrics to evaluate the accuracy of our predictions - the Dice coefficient, 95th percentile Hausdorff distance, and average surface distance (ASD). These metrics are implemented using the SimpleITK Python package \cite{simpleitk1, simpleitk2, simpleitk3}. We briefly describe these metrics below.

\emph{Dice Coefficient} - As described in the Dice loss function description, the Dice coefficient is a widely used measure of overlap in computer vision tasks. More specifically, it measures the overlap between the two binary sets. It ranges from 0 to 1, where 1 indicates a perfect match between the predicted and ground truth segmentations.

\emph{95th Percentile Hausdorff Distance} - While the Dice score is a commonly used metric for comparing two segmentation masks, it is not sensitive to local differences, as it represents a global measure of overlap. Therefore, we compute a complementary metric, the 95th percentile Hausdorff distance (HD95), which is a distance metric that measures the maximum of the minimum distances between the predicted segmentation and the ground truth at the 95th percentile. The HD95 is a non-negative real number measured in millimeters, with a value of 0mm indicating a perfect prediction.

\emph{Average Surface Distance} - The average surface distance (ASD) measures the average distance between the predicted segmentation and the ground truth along the surface of the object being segmented. It is a more fine-grained metric than the HD95 since it captures the surface-level details of the segmentation and can provide insight into the quality of the segmentation. The ASD is a non-negative real number measured in millimeters with a perfect prediction achieving a value of 0mm.

\section{Results}
\label{sec:results}
\begin{figure*}
    \centering
    \includegraphics[width=\textwidth]{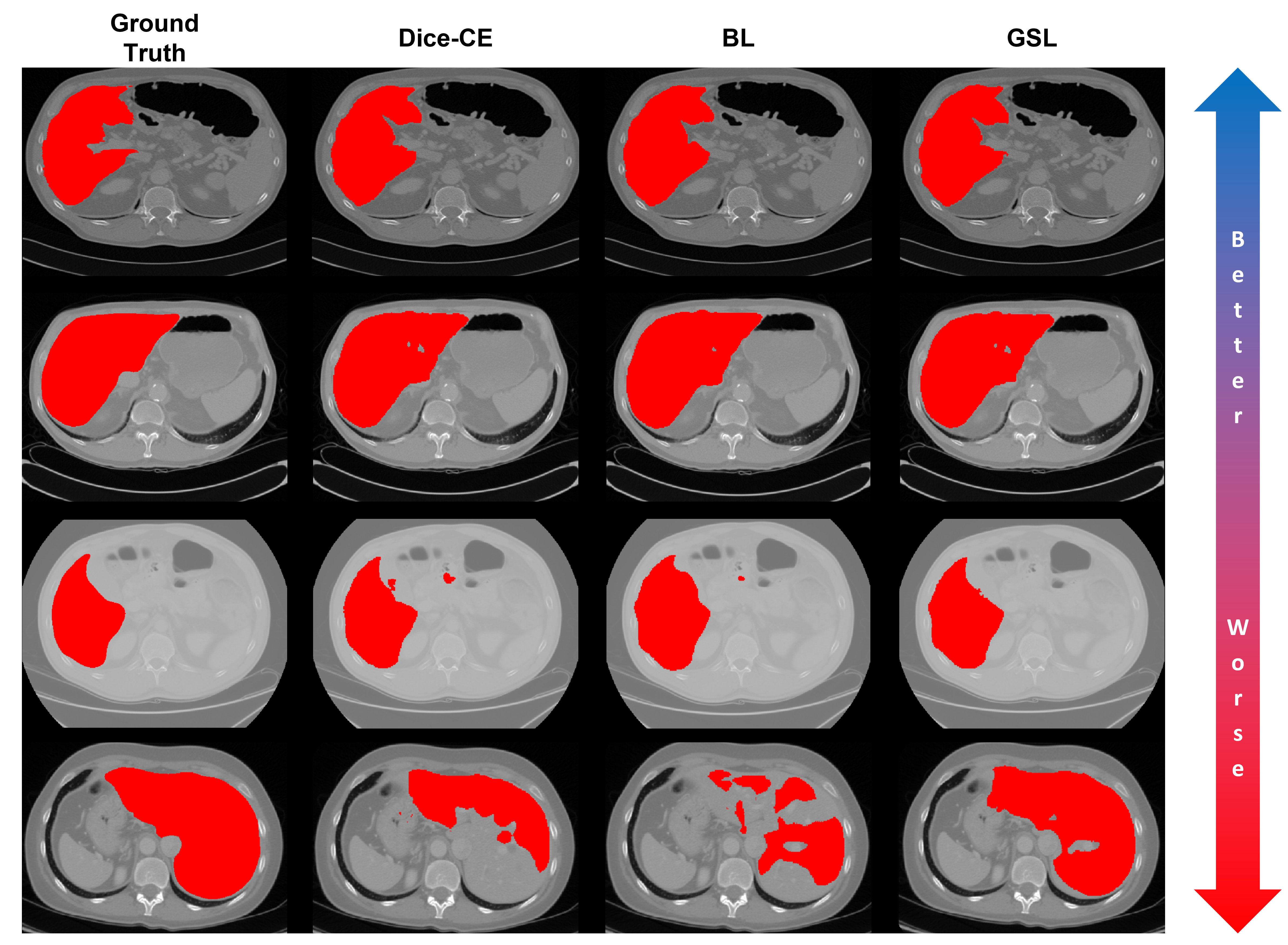}
    \caption{From left to right, ground truth and predictions from the nnUNet architecture trained on LiTS data with Dice-CE, BL, and GSL functions for a spectrum of easier to more difficult test cases. Here, we see that, even for more difficult cases, the GSL produces visually superior predictions than the Dice-CE and BL functions. \label{fig:lits-preds}}
\end{figure*}

\begin{figure*}
    \centering
    \includegraphics[width=\textwidth]{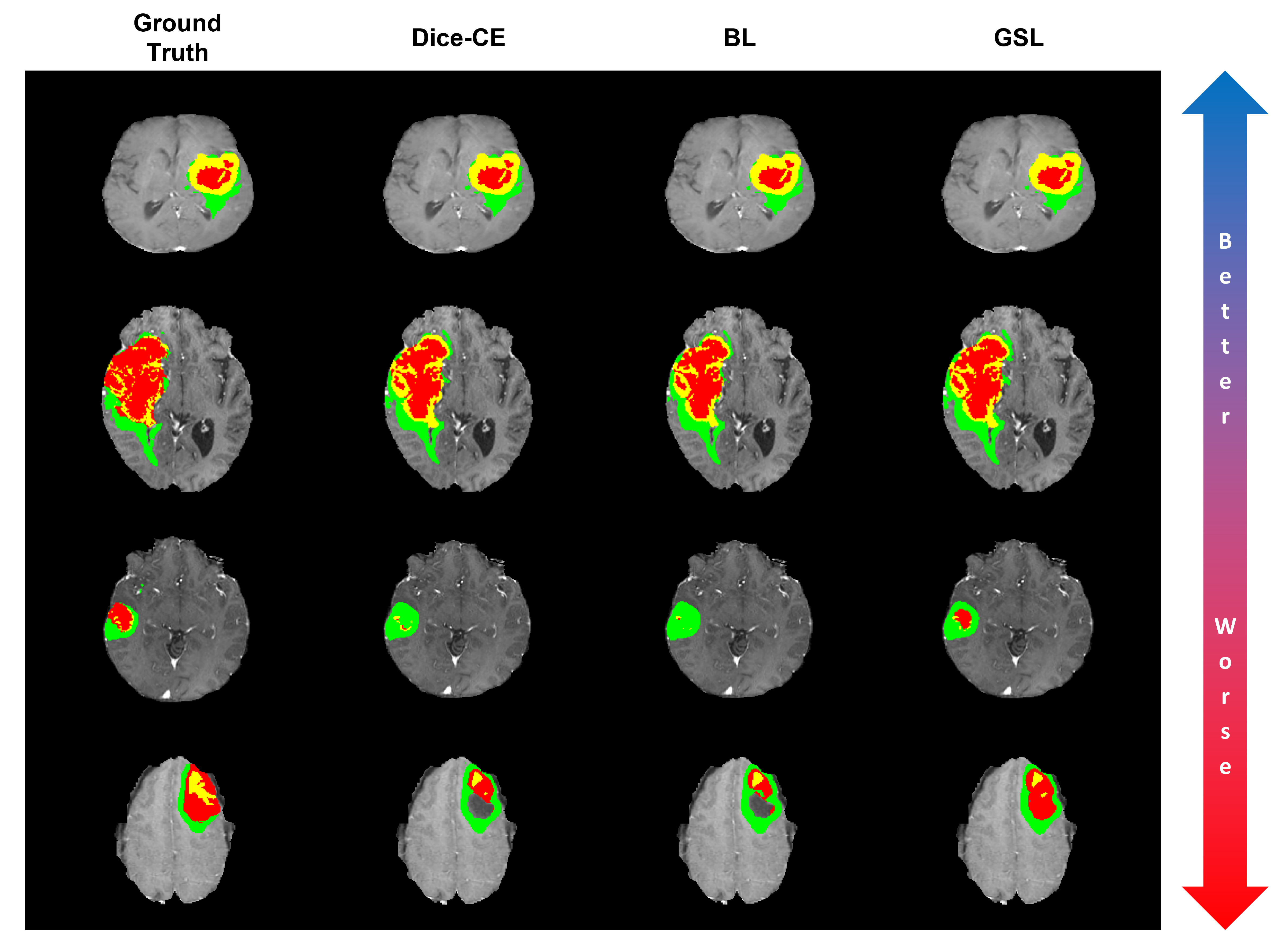}
    \caption{From left to right, ground truth and predictions from the nnUNet architecture trained on BraTS data with Dice-CE, BL, and GSL functions for a spectrum of easier to more difficult test cases. Here, we see that, even for more difficult cases, the GSL produces visually superior predictions than the Dice-CE and BL functions. \label{fig:brats-preds}}
\end{figure*}

Using the methods described in Section \ref{sec:methods}, we train a nnUNet using each loss function described in Section \ref{sec:prevwork} and compare their performance to our proposed GSL function. This experiment uses a linear schedule for the parameter $\alpha$ in the boundary-based losses (i.e., HL, BL, and GSL). Table \ref{tab:results-main} shows the results of this comparison. Here, our GSL achieves lower Hausdorff 95 and average surface distances for the LiTS and BraTS challenge datasets. Figures \ref{fig:lits-preds} and \ref{fig:brats-preds} show from left to right the ground truth and predictions from the nnUNet architecture trained on LiTS and BraTS data respectively, with the Dice-CE, BL, and GSL functions for a spectrum of easier to more difficult test cases. Even for more difficult cases, we see that the GSL produces visually superior predictions than the Dice-CE and BL functions. 

\begin{table*}[ht!]
\centering
\bgroup
\def\arraystretch{1.15}
\resizebox{0.75\textwidth}{!}{%
\begin{tabular}{cclccc}
\hline
Dataset & Task & Loss & Dice [$\uparrow$] & Hausdorf 95 (mm) [$\downarrow$] & Avg. Surface (mm) [$\downarrow$] \\ \hline
\multirow{6}{*}{LiTS} & \multirow{6}{*}{Liver} & DL & \textbf{0.9304 (0.0974)} & 9.0655 (30.530) & 3.7608 (11.976) \\
 &  & Dice-CE & 0.9287 (0.1229) & 10.874 (35.205) & 3.7455 (11.361) \\
 &  & GDL & 0.9246 (0.1317) & 9.0825 (28.023) & 3.6259 (10.693) \\
 &  & HL & 0.9166 (0.1237) & 17.451 (71.162) & 9.5963 (64.328) \\
 &  & BL & 0.9270 (0.1221) & 12.366 (66.451) & 7.9931 (63.268) \\
 &  & GSL & 0.9302 (0.1075) & \textbf{8.9046 (28.205)} & \textbf{3.2791 (7.0823)} \\ \hline
\multirow{18}{*}{BraTS} & \multirow{6}{*}{Whole Tumor} & DL & 0.9027 (0.0784) & 3.9117 (10.236) & 1.2931 (2.0569) \\
 &  & Dice-CE & 0.9048 (0.0771) & 4.2284 (12.342) & 1.2852 (2.0275) \\
 &  & GDL & 0.8508 (0.1303) & 7.6853 (16.766) & 2.3204 (3.5123) \\
 &  & HL & 0.8607 (0.1117) & 5.0997 (10.997) & 1.7400 (2.2935) \\
 &  & BL & 0.9011 (0.0740) & 3.9045 (10.036) & 1.2400 (1.7657) \\
 &  & GSL & \textbf{0.9087 (0.0722)} & \textbf{3.5367 (9.7121)} & \textbf{1.1442 (1.8159)} \\ \cline{2-6} 
 & \multirow{6}{*}{Tumor Core} & DL & 0.8403 (0.1803) & 5.5020 (12.859) & 1.7861 (3.7423) \\
 &  & Dice-CE & 0.8413 (0.1742) & 5.8953 (15.191) & 1.7464 (3.1522) \\
 &  & GDL & 0.7799 (0.2640) & 7.7837 (16.793) & 2.6613 (5.1596) \\
 &  & HL & 0.7519 (0.2102) & 7.8241 (21.669) & 3.5736 (19.697) \\
 &  & BL & 0.8318 (0.1781) & 5.4368 (12.532) & \textbf{1.7446 (3.2988)} \\
 &  & GSL & \textbf{0.8448 (0.1835)} & \textbf{4.9526 (11.333)} & 1.8121 (4.9699) \\ \cline{2-6} 
 & \multirow{6}{*}{Enhancing Tumor} & DL & 0.7422 (0.2791) & 32.704 (96.645) & 28.640 (96.886) \\
 &  & Dice-CE & 0.7424 (0.2773) & 32.593 (96.689) & 28.539 (96.878) \\
 &  & GDL & 0.7431 (0.2738) & 32.369 (95.088) & 27.792 (95.176) \\
 &  & HL & 0.6831 (0.2773) & 32.196 (94.416) & 27.833 (95.141) \\
 &  & BL & 0.7395 (0.2770) & 30.971 (94.723) & 27.520 (95.218) \\
 &  & GSL & \textbf{0.7587 (0.2696)} & \textbf{30.121 (93.228)} & \textbf{26.551 (93.568)} \\ \hline
\end{tabular}%
}
\egroup
\caption{Mean and standard deviation of each metric for each loss function using the nnUNet architecture. Here, we see that our GSL achieves lower Hausdorff 95 and average surface distances for the LiTS and BraTS challenge datasets.}
\label{tab:results-main}
\end{table*}

We also test our GSL with different $\alpha$-schedules. Table \ref{tab:results-schedule} shows each metric's mean and standard deviation from testing different schedules for $\alpha$ with our GSL. We compare the accuracy for linear, decreasing step functions with step lengths of 5, 25, and 50 epochs and a cosine function as schedules. For the LiTS dataset, it is clear that the decreasing step function with a step length of 5 epochs achieves the best accuracy vs. the other schedules. For the BraTS dataset, the results are less clear, with the decreasing step function with a step length of 25 epochs achieving the best metrics for the Whole Tumor task, the step function with a step length of 5 epochs for the Tumor Core task, and the linear schedule for the Enhancing Tumor task.

\begin{table*}[ht!]
\centering
\bgroup
\def\arraystretch{1.15}
\resizebox{0.75\textwidth}{!}{%
\begin{tabular}{cclccc}
\hline
Dataset & Task & \multicolumn{1}{c}{Schedule} & Dice [$\uparrow$] & Hausdorf 95 (mm) [$\downarrow$] & Avg. Surface (mm) [$\downarrow$] \\ \hline
\multirow{5}{*}{LiTS} & \multirow{5}{*}{Liver} & Linear & 0.9302 (0.1075) & 8.9046 (28.205) & 3.2791 (7.0823) \\
 &  & Step - 5 & \textbf{0.9339 (0.0983)} & \textbf{7.2410 (24.915)} & \textbf{2.6315 (5.0905)} \\
 &  & Step - 25 & 0.9302 (0.1133) & 8.0823 (26.251) & 3.1676 (8.5182) \\
 &  & Step - 50 & 0.9275 (0.1107) & 7.9524 (24.056) & 3.1763 (6.0171) \\
 &  & Cosine & 0.9300 (0.1182) & 13.005 (67.511) & 8.2237 (63.309) \\ \hline
\multirow{15}{*}{BraTS} & \multirow{5}{*}{Whole Tumor} & Linear & 0.9087 (0.0722) & 3.5367 (9.7121) & 1.1442 (1.8159) \\
 &  & Step - 5 & 0.9081 (0.0738) & 3.7796 (9.4055) & 1.1552 (1.6483) \\
 &  & Step - 25 & \textbf{0.9093 (0.0786)} & \textbf{2.9509 (8.4327)} & \textbf{1.0719 (1.7167)} \\
 &  & Step - 50 & 0.9067 (0.0775) & 3.8417 (10.776) & 1.2326 (2.1089) \\
 &  & Cosine & 0.9074 (0.0833) & 3.6255 (10.140) & 1.2187 (2.6784) \\ \cline{2-6} 
 & \multirow{5}{*}{Tumor Core} & Linear & 0.8448 (0.1835) & 4.9526 (11.333) & 1.8121 (4.9699) \\
 &  & Step - 5 & 0.8439 (0.1804) & \textbf{4.8878 (10.846)} & \textbf{1.7734 (4.6574)} \\
 &  & Step - 25 & 0.8473 (0.1795) & 6.5636 (28.895) & 3.5387 (27.425) \\
 &  & Step - 50 & 0.8473 (0.1776) & 5.5879 (21.672) & 2.4524 (19.434) \\
 &  & Cosine & \textbf{0.8508 (0.1750)} & 5.3815 (21.941) & 2.6191 (19.833) \\ \cline{2-6} 
 & \multirow{5}{*}{Enhancing Tumor} & Linear & \textbf{0.7587 (0.2696)} & \textbf{30.121 (93.228)} & \textbf{26.551 (93.568)} \\
 &  & Step - 5 & 0.7517 (0.2759) & 31.159 (94.948) & 27.521 (95.237) \\
 &  & Step - 25 & 0.7555 (0.2739) & 32.453 (98.038) & 29.318 (98.554) \\
 &  & Step - 50 & 0.7511 (0.2781) & 31.863 (96.466) & 28.352 (96.909) \\
 &  & Cosine & 0.7562 (0.2707) & 31.663 (96.562) & 28.352 (96.918) \\ \hline
\end{tabular}%
}
\egroup
\caption{Each metric's mean and standard deviation from testing different schedules for $\alpha$ with our GSL. We compare the accuracy for linear, decreasing step functions with step lengths of 5, 25, and 50 epochs, and a cosine function as schedules. For the LiTS dataset, the decreasing step function with a step length of 5 epochs achieves the best accuracy vs. the other schedules. For the BraTS dataset, the results are less clear, with the decreasing step function with a step length of 25 epochs achieving the best metrics for the Whole Tumor task, the step function with a step length of 5 epochs for the Tumor Core task, and the linear schedule for the Enhancing Tumor task.}
\label{tab:results-schedule}
\end{table*}

\section{Discussion}
The results above show that our proposed GSL function is a promising alternative loss function for medical imaging segmentation tasks. The GSL outperforms other losses regarding HD and ASD accuracy on the LiTS and BraTS datasets while maintaining comparable accuracy for the Dice coefficient. The GSL predictions also generally appear to have less variance than the other losses tested. These results also suggest that the GSL function could benefit applications where HD and ASD accuracy are crucial. We hypothesize that our proposed loss function archives higher accuracy than other boundary-based loss functions partly because it is normalized to a similar scale as the region-based loss with which it is used in a weighted combination. This normalization is desirable from an optimization perspective and a desirable property for training in the presence of noise. Indeed, \cite{braun2020data} and \cite{ma2020normalized} show that normalized loss functions can improve the robustness of deep learning models with noise in the data or labels. Medical imaging data is inherently noisy, with CT images showing a degree of Gaussian noise and MR images showing Rician noise and bias fields, depending on the machine and acquisition parameters \cite{diwakar2018review, mayasari2019reduce, coupe2010robust, sing2015estimation}.

The results in Table \ref{tab:results-schedule} indicate that the choice of scheduler for the weighting parameter $\alpha$ in (\ref{eqn:scheduled-loss}). For the LiTS dataset, the choice of a decreasing step function with a step length of 5 epochs appears to be the optimal choice vs. the other schedules. However, the results are unclear for the BraTS dataset, with the linear and step functions with various step length sizes appearing to produce the most accurate result for different metrics and tumor subcomponents. The intuition governing the step function schedules is to allow the optimizer to focus on fewer subproblems during training. One can view the linear function as a decreasing step function with step length one. However, increasing the size of the step length allows optimizers like Adam to minimize each stage of the overall loss more effectively. Our results partially support this claim, but further work is needed to understand better how to select an $\alpha$-schedule optimally.

The intuition behind the GSL function is based on the properties of the segmentation images (values are in the interval $[0, 1]$) and the DTM. Recall that for a given segmentation, the DTM is positive on the exterior, zero on the boundary, and negative on the object's interior. Hence, for the given maximal value in (\ref{eqn:worst}), the goal of the GLS is to produce a prediction $P$ such that $D \odot (\mathbf{1} - (T + P)) = |D|$, where $\odot$ is the Hadamard or pointwise product. In other words, predictions that are as close as possible to the ground truth will also recover the absolute value of the DTM. Note that we use the 2-norm instead of absolute values in our formulation. Using other norms like the 1-norm may affect the results presented above. Future work will explore developing and testing different formulations of the GSL.

While not tested in our work, the choice of region-based loss for (\ref{eqn:scheduled-loss}) may also affect the results shown in Tables \ref{tab:results-main} and \ref{tab:results-schedule}. In several cases in Table \ref{tab:results-main}, we see that the Dice loss achieves higher accuracy than the Dice-CE loss. We may improve our results by using the Dice loss as $\mathcal{L}_{region}$. Additionally, using the precomputed weights shown in (\ref{eqn:weights}) for the GDL may also serve as an effective region-based loss. It may also be worth considering non-region-based losses in (\ref{eqn:scheduled-loss}). For example, using only cross-entropy or the focal loss \cite{lin2017focal}, which are considered distribution-based losses \cite{loss-survey}, could be beneficial. Testing different region (or non-region) losses will be an objective of future work.

The choice of precomputed weighting terms may be another important factor in our GSL. For example, one might also consider the weighting terms
\begin{align}
    w_k = \left( \frac{1}{\sum_{j=1}^C \left( \frac{1}{N_j}\right)^p} \right) \left( \frac{1}{N_k}\right)^p,
\end{align}
for some $p > 1$. Additionally, weighting terms based on the surface area of the given object might be advantageous for the GSL. In \cite{sugino2021loss}, Sugino et al. show the effectiveness for weights that depend on the DTM itself. It is also unclear how the weighting terms and the given segmentation task are related. For example, other weighting schemes might be more appropriate for different types of tumors or imaging modalities. 

\section{Conclusion}
Our results indicate that we can improve segmentation accuracy for deep learning-based medical imaging segmentation tasks with our novel GSL function. When tested on the BraTS and LiTS datasets, the state-of-the-art nnUNet architecture trained with our proposed GSL achieved greater accuracy in the HD and ASD and comparable Dice scores. While further testing on other datasets like the Medical Segmentation Decathalon \cite{antonelli2022medical} is needed, we are encouraged by the level of accuracy observed in the BraTS and LiTS datasets. Future work will focus on continuing to validate our results on more diverse and complex segmentation tasks and further refining our GLS to incorporate better $\alpha$-schedules, more refined weighting terms, and optimal complementary losses in (\ref{eqn:scheduled-loss}).

\section{Acknowledgments}
The Department of Defense supports Adrian Celaya through the National Defense Science \& Engineering Graduate Fellowship Program. David Fuentes is partially supported by R21CA249373. Beatrice Riviere is partially supported by NSF-DMS2111459. This research was partially supported by the Tumor Measurement Initiative through the MD Anderson Strategic Research Initiative Development (STRIDE), NSF-2111147, and NSF-2111459.

{\small
\bibliographystyle{ieee_fullname}
\bibliography{sources}
}

\end{document}